\documentclass{article}[11pt]
\usepackage{a4wide}
\usepackage{stmaryrd,latexsym,bm,bbm}
\usepackage{epsfig,graphics}
\usepackage{rotfloat}
\usepackage{amsmath,amssymb,amsfonts}
\usepackage{color}
\usepackage{epstopdf}
\usepackage{slashed}

\usepackage[font=footnotesize, labelfont=bf]{caption,subcaption}
\usepackage[square,comma,sort&compress,numbers]{natbib}
\usepackage[
colorlinks=true,
linkcolor=black,
breaklinks=true,
urlcolor=blue,
citecolor=green]{hyperref}
\usepackage[utf8]{inputenc}

\usepackage{booktabs}
\usepackage{dcolumn}
\usepackage{makecell}
\usepackage{multirow}

\definecolor{darkblue}{rgb}{0.,0.,0.7}
\definecolor{light-blue}{rgb}{0.8,0.85,1}
\definecolor{green}{rgb}{0,0.6,0}
\definecolor{blueviolet}{rgb}{0.541, 0.169, 0.886}
\definecolor{fuchsia}{rgb}{1.0, 0, 1.0}

\DeclareMathOperator{\dif}{d\!}
\newcommand{\order}[1]{\mathcal{O}\left(#1\right)}

\newcommand{\abs}[1]{\lvert#1\rvert}

\newcommand{\mmev}{\mathrm{meV}}

\newcommand{\ket}[1]{\left| #1 \right\rangle}
\newcommand{\bra}[1]{\left\langle #1 \right|}
\newcommand{\ketdot}[2]{\langle #1 | #2\rangle}

\newcommand{\beq}{\begin{equation}}
\newcommand{\eeq}{\end{equation}}
\newcommand{\beqa}{\begin{eqnarray}}
\newcommand{\eeqa}{\end{eqnarray}}

\begin{document}

\title{Nuclear deformation effects on charge radius measurements of the proton and deuteron }
\date{\today}
\author{Yong-Hui Lin$^{1,2,}$\footnote{Email address:
      \texttt{linyonghui@itp.ac.cn} }~ and
      Bing-Song Zou$^{1,2,3,}$\footnote{Email address:
      \texttt{zoubs@itp.ac.cn} }
        \\[2mm]
      {\it\small$^1$CAS Key Laboratory of Theoretical Physics, Institute
      of Theoretical Physics,}\\
      {\it\small  Chinese Academy of Sciences, Beijing 100190,China}\\
      {\it\small$^2$ School of Physical Sciences, University of Chinese Academy
of Sciences, Beijing 100049, China} \\
{\it\small$^3$ School of Physics, Central South University, Changsha 410083, China} 
}

\maketitle

\begin{abstract}Up to now, all charge radius measurements of the proton and deuteron assumed uniform spheroidal charge distribution.
We investigate the nuclear deformation effects on these charge radius measurements by assuming a uniform prolate charge distribution for the proton and deuteron. We solve the energy levels of the corresponding muonic and electric atoms with such deformed nucleus and present how the purely quadruple deformation of proton and deuteron affects their Lamb shifts. The numerical results suggest that the deformation of proton and deuteron leads to that the charge radius extracted from the electronic measurement should be smaller than the corresponding one in the muonic measurement which assumed uniform spheroidal charge distribution. If the central values of newest measurements for the proton are adopted, the proton would have a prolate structure with the 0.91 $\mathrm{fm}$ long axis and 0.73 $\mathrm{fm}$ short axis. Further improved precise charge radius measurements of the proton and deuteron will help us to pin down their shape deformation. 

\end{abstract}

\medskip
\newpage

\section{Introduction} \label{sec:1}
In 2010, an experimental measurement of the proton charge radius with the unprecedented precision caused a great stir at that time. This experiment was performed by the CREMA collaboration. The proton electric charge radius $r_E^p = 0.84087(39)~\mathrm{fm}$ was deduced from the accurate measurement of the Lamb shift in muonic hydrogen~\cite{Pohl:2010zza}. This surprising result shows a $5.6\sigma$ discrepancy with the previously accepted radius $r_E^p = 0.8751(61)~\mathrm{fm}$ which measured from a combination of ordinary hydrogen spectroscopy and electron-proton scattering~\cite{Mohr:2015ccw}, which is known as proton radius puzzle. Six years later, the similar discrepancy in the measurements of deuteron charge radius was also reported by Ref.~\cite{Pohl1:2016xoo}.  A lot of attempts from both theoretical and experimental sides have been made to understand this puzzle~\cite{Pohl:2013yb,Hill:2017wzi}.
Within past two years, two additional measurements in hydrogen have been published to determine the proton radius: a measurement of the $2S\to 4P$ interval~\cite{Beyer:2017gug} and a measurement of the $1S\to 3S$ interval~\cite{Fleurbaey:2018fih}. While Ref.~\cite{Beyer:2017gug} yielded a small radius $r_p=0.8335(95)~\mathrm{fm}$ in consistent with the result from muonic hydrogen measurement, Ref.~\cite{Fleurbaey:2018fih} gave a much larger radius $r_p=0.877(13)~\mathrm{fm}$ in consistent with the old averaged value before 2010.
Very recently, in this year, Bezginov et al.~\cite{Bezginov:2019mdi} measured the Lamb shift in electronic hydrogen, which allowed for a direct comparison to the Lamb shift measured in muonic hydrogen.  The new measurement determines the proton radius to be $r_p=0.833(10)~\mathrm{fm}$
supporting the smaller value of Refs.~\cite{Pohl:2010zza,Beyer:2017gug}. 

The most noticeable fact about the proton radius puzzle is that the muon mass is about 200 times the electron mass, that is $m_{\mu}/m_e \sim 200$. It implies that the size of muonic hydrogen is around 200 times smaller than electronic hydrogen. And the energy levels of muonic hydrogen are much more sensitive to the proton structure than electronic hydrogen. Thus, the correction from the proton structure will play an important role when extracting the proton radius from the frequency measurement of the 2P-2S transition in muonic atoms. Inspired by this reality, the proton-structure correction to the hyperfine splitting in the light hydrogen-like atoms have been systematically estimated in Refs.~\cite{Friar:1978wv,Eides:2000xc,Antognini:2013jkc,Jentschura:2010ej,Carlson:2011zd,Repko:2017pcl,Pachucki:1999zza,Borie:2004fv,Ji:2018ozm,Krauth:2015nja}. However all the results in these references show that the correction to the Lamb shift of muonic hydrogen from the QED effects even including the high-order contribution($\sim \order{\alpha^6}$) is not large enough to solve the proton radius puzzle. 

Up to now, all charge radius measurements of the proton and deuteron and relevant theoretical studies assumed uniform spheroidal charge distribution. But we know that due to a mixture of D-wave component of two nucleons in the deuteron, the shape of deuteron is deformed. 
The existence of an intrinsic deformation for nucleon has also been demonstrated by some early works~\cite{Buchmann:2001gj,Bernstein:2002fe,Buchmann:2005mp,Buchmann:2007zz,Vanderhaeghen:2010nd}. In the present work, we analyze the behaviors of energy levels both in muonic and electronic atoms with the uniform prolate charge distribution assumption for the proton and deuteron. 

\section{Formalism} \label{sec:2}
Due to the non-perturbative nature of QCD at low energy scale, we cannot derive the charge density of proton from the first principle. In this work, we describe the non-spherical structure of proton with a uniformly prolate charge distribution which has been suggested in the Refs.~\cite{Buchmann:2001gj,Bernstein:2002fe,Buchmann:2005mp,Buchmann:2007zz} and adopt the \emph{geometric collective model} to parameterize the deformed proton surface. For the case of the pure quadrupole deformation the nuclear surface is described as
\begin{equation}\label{eq:surface}
R(\theta)=r_0 \left( 1+\frac{1}{4} \sqrt{\frac{5}{4\pi}} \beta (1+3\cos(2\theta)) \right),
\end{equation}
where $\beta$ is the deformation parameter, $\beta>0$ for the prolate shape with two same short axes along the x, y-axis directions and a long axis along the z axis and $\beta<0$ for the oblate shape with two same long axes along the x, y-axis directions and a short axis along the z axis. The corresponding charge density can be read as
\begin{equation}\label{eq:density}
\rho(r,\theta)=\frac{3/\left(4\pi r_0^3\right)}{1+\exp\left(\frac{r-R(\theta)}{a} \right) }.
\end{equation}
Actually, the density for the uniformly prolate charge distribution should be a piecewise function and here we apply the Woods-Saxon approximation to convert the piecewise-like expression of the charge density to the continuous one. The free parameter $a$  should be a small value close to zero. And the numerator of Eq.~\eqref{eq:density} is the charge density constant inside the proton. Furthermore, the electric potential derived by this deformed charge density of proton can be obtained as following,
\begin{equation}
\phi(\bm r)=\int\dif v^\prime\frac{\rho(\bm r^\prime)}{\abs{\bm r-\bm r^\prime}}.
\end{equation}
Expanding the Green's function in terms of spherical harmonics,
\begin{equation}
\frac{1}{\abs{\bm r-\bm r^\prime}}=4\pi\sum_{\lambda=0}^{\infty}\sum_{\mu=-\lambda}^{\lambda}\frac1{2\lambda+1}\frac{r^\lambda_{<}}{r^{\lambda+1}_{>}}Y^*_{\lambda\mu}(\theta,\phi)Y_{\lambda\mu}(\theta^\prime,\phi^\prime),
\end{equation}
where $r_<$ and $r_>$ stand for the smaller and larger one of the two radii $r$ and $r^\prime$, respectively. For the system with axial symmetry, it gets
\begin{equation}
\frac{1}{\abs{\bm r-\bm r^\prime}}=\sum_{\lambda=0}^{\infty}\frac{r^\lambda_{<}}{r^{\lambda+1}_{>}}P_{\lambda}(\cos\theta)P_{\lambda}(\cos\theta^\prime).
\end{equation}
Then the electrostatic potential becomes
\begin{equation}\label{potential}
\phi(r,\theta)=2\pi\sum_{\lambda}P_\lambda(\cos{\theta})\left(\frac{1}{r^{\lambda+1}}\int_0^r\dif r^\prime r^{\prime\lambda} \rho_\lambda(r^\prime)+r^\lambda\int_r^\infty \dif r^\prime \frac{1}{r^{\prime\lambda+1}}\rho_\lambda(r^\prime)\right),
\end{equation}
with
\begin{equation}
\rho_\lambda(r^\prime)=\int\dif\cos{\theta^\prime}\rho(r^\prime,\theta^\prime)P_\lambda(\cos{\theta^\prime}).
\end{equation}
For our quadrupole deformation assumption, the above summation of partial wave is taken up to $\lambda=2$. After calculating the potential, we come to consider the dynamics of the electronic and muonic atoms. Before going forward, it must be kept in mind that our purpose is to estimate the correction of the nonspherical charge distribution effect to the Lamb shift. Actually, as we can see from Ref.~\cite{Eides:2000xc}, the leading order nuclear size correction($\sim \order{\alpha^4}$) is the dominating part of the Lamb shift which caused by the nuclear non-point effects. And all other high order corrections are several orders of magnitude smaller. Considering the desired $0.31\ \mathrm{meV}$ energy shift for compensating the proton radius puzzle, it is safe to ignore the nonspherical charge distribution effect to high order contribution. Note that the leading order nuclear size correction which is proportional to the root-mean-square(RMS) radius of proton~$\left\langle r_p^2 \right\rangle $ can be reproduced by considering the uniform spheroidal charge distribution with an effective radius $r_0=c\sqrt{\left\langle r_p^2\right\rangle}$ for the proton. Here $c$ is a constant that can be determined by fitting the energy shift between $2S$ and $2P$ levels of hydrogen to the calculation of QED. After the ratio $c$ obtained, what we need to do is that calculating the energy shift between $2S$ and $2P$ states with the uniformly prolate charge distribution and subtracting the part from the uniform spheroidal potential. Then the obtained remainder is the correction to the lamb shift which we want to issue.

The Hamiltonian for the two-body electromagnetic system states
\begin{equation}
H=-\frac{\hbar^2}{2\mu}\nabla^2+V(\bm r),
\end{equation}
where $\mu$ is the reduced mass and $V(\bm r)=-e\phi(\bm r)$ is the electromagnetic potential. Then the equation of motion is read as
\begin{equation}
\left[-\frac{\hbar^2}{2\mu}\nabla^2+V(\bm r)-E \right]\psi(\bm r)=0.
\end{equation}
In order to solve this eigenvalue problem, the variational principle is used and we adopt the Gaussian Expansion Method proposed in Ref.~\cite{Hiyama:2003cu}. To be concrete, the wave function $\psi(\bm r)$ is expanded in terms of a set of Gaussian basis functions:
\begin{align}
&\psi(\bm r)=\sum_{n=1}^{n_{max}}c_{nl}\phi_{nlm}^G(\bm r),\\
&\phi_{nlm}^G(\bm r)=\phi_{nl}^G(r)Y_{lm}(\hat{\bm r}),\\
&\phi_{nl}^G(r)=N_{nl}r^l e^{-\nu_n r^2},\\
&N_{nl}=\left(\frac{2^{l+2}(2\nu_n)^{l+\frac32}}{\sqrt{\pi}(2l+1)!!}\right)^{\frac12}\quad (n=1,2,3,\cdots,n_{max}).
\end{align}
The constant $N_{nl}$ is determined from the normalization $\ketdot{\phi_{nlm}^G}{\phi_{nlm}^G}=1$. As Ref.~\cite{Hiyama:2003cu} suggested, the best set of Gaussian size parameters can be choosed as
\begin{align}
\nu_n&=\frac{1}{r_n^2},\notag\\
r_n&=r_1 a^{n-1}\quad (n=1,2,3,\cdots,n_{max}).
\end{align}
There are three parameters, $\{n_{max}, r_1, r_{n_{max}}\}$ or $\{n_{max}, r_1, a\}$ of which we use the former type in this work. The expansion coefficients $\{c_{nl}\}$ and the eigenenergy $E$ are determined by the Rayleigh-Ritz variational principle, which leads to a generalized matrix eigenvalue problem,
\begin{equation}\label{eq:eigenvalue}
\sum_{p'=p_1}^{p_k}\left[(T_{pp'}+V_{pp'})-EN_{pp'}\right]c_{p'}=0,\qquad(p=p_1,p_2,p_3,\cdots,p_k),
\end{equation}
where the matrix elements are given by
\begin{align}
N_{pp'}&=\ketdot{\phi_{nlm}^G}{\phi_{n'l'm'}^G}=\left(\frac{2\sqrt{\nu_n\nu_{n'}}}{\nu_n+\nu_{n'}}\right)^{l+\frac32}\delta_{ll'}\delta_{mm'},\\
T_{pp'}&=\bra{\phi_{nlm}^G}-\frac{\hbar^2}{2\mu}\nabla^2\ket{\phi_{n'l'm'}^G}=\frac{\hbar^2}{\mu}\frac{(2l+3)\nu_n\nu_{n'}}{\nu_n+\nu_{n'}}\left(\frac{2\sqrt{\nu_n\nu_{n'}}}{\nu_n+\nu_{n'}}\right)^{l+\frac32}\delta_{ll'}\delta_{mm'},\\
V_{pp'}&=\bra{\phi_{nlm}^G}V(\bm r)\ket{\phi_{n'l'm'}^G},
\end{align}
and $p_i$ denotes the quantum numbers of the $i$th Gaussian basis function, that is $\{n_i,l_i,m_i\}$. The summation in the Eq.~\eqref{eq:eigenvalue} is over all the Gaussian basis functions. The number of expanded basis function, $k$, equals $n_{max}$ for the spherical symmetry potential, while $k=n_{max} (n_{max}+1)/2$ for the potential with axial symmetry.

\section{Numerical results} \label{sec:3}
For the extraction of a single radius parameter of proton from the experimental hydrogen atom spectroscopy, it is useful to write the lamb shift in the form
\begin{equation}\label{eq:lambshift}
	\Delta E_{LS}=\Delta E_{radius-independent}+\Delta E_{radius-dependent}.
\end{equation}
The radius-dependent contribution can be parameterized customarily as $A\left\langle r_p^2 \right\rangle+B(\left\langle r_p^2 \right\rangle)^{3/2}$ for the muonic system and $A\left\langle r_p^2 \right\rangle$ for the electronic system. And the radius-independent terms collect all the other QED contributions. The coefficient $A$ can be determined straightforwardly: Up to order of $\alpha(Z\alpha)^5$, there are two contributions, that is, the leading nuclear size corrections~($\sim \order{(Z\alpha)^4}$) and the radiative correction of order $\alpha(Z\alpha)^5$ to it~\cite{Eides:2000xc}. The former term can be written as
\begin{align}\label{eq:finiteness}
E_{FS}&=\bra{nS}\delta V\ket{nS}=\bra{nS}-\left.\frac{\dif G_E}{\dif Q^2}\right|_{Q^2=0}\Delta V \ket{nS}\notag\\
&=\frac{1}{6}\left\langle r^2\right\rangle 4\pi(Z\alpha)\bra{nS}\delta(\bm r) \ket{nS}=\frac{2\pi(Z\alpha)}{3}\left\langle r^2\right\rangle\abs{\psi(0)}^2\notag\\
&=\frac{2\pi(Z\alpha)}{3}\left\langle r^2\right\rangle\frac{1}{\pi}\left(\frac{Z\alpha m_r}{n}\right)^3\delta_{l0}=\frac{2(Z\alpha)^4}{3n^3}m_r^3\left\langle r^2\right\rangle\delta_{l0},
\end{align}
where $m_r$ is the reduced mass and $Z$ denotes the charge number of nucleus. Note that $\left\langle r^2\right\rangle=-6\left.\frac{\dif G_E}{\dif Q^2}\right|_{Q^2=0}$ is obtained by assuming spherical symmetry of charge density for the nucleus~\cite{Pacetti:2018wwk}. As shown in Ref.~\cite{Pachucki:1999zza}, the value of latter term is about the magnitude of the order of $0.01\ \mmev$ which is negligible in the present work. However, the Zemach-moment term $B(\left\langle r_p^2 \right\rangle)^{3/2}$ indicates the contribution of typically two-photon exchanges. The value of coefficient $B$ depends on a model for the form factor of the nucleus~\cite{Borie:2004fv}. And it can be ascertained roughly that the magnitude of this term  is about on the order of $m(Z\alpha)^5$~\cite{Friar:1978wv}. Since the small mass of electron, the Zemach-moment term is dropped usually in the electronic system. Finally, we collect the theoretical expressions of $\Delta E_{radius-dependent}$ which are used customarily to extract the charge radius of proton and deuteron from the experimental spectroscopy data in Table~\ref{table:finiteness}.
\begin{table}[htpb]
	\centering
	\caption{\label{table:finiteness}The theoretical formula for the radius-dependent part of the lamb shift. $r_{p,e}$ and $r_{p,m}$ denote the $\sqrt{\left\langle r^2\right\rangle}$ of proton which is extracted from the electronic and muonic hydrogen spectroscopy,respectively. And for the deuteron, we adopt the similar symbols $r_{d,e}$ and $r_{d,m}$. The energy shifts $\Delta E$ are in unit of $\mathrm{meV}$ and the charge radius $r_{p(d),e}$ and $r_{p(d),m}$ are in unit of $\mathrm{fm}$.}
	\scalebox{1.0}{
		\begin{tabular}{*{11}{c}}
			\toprule\morecmidrules\toprule
			$\Delta E_{radius-dependent}$ & Measurements	\\
			\Xhline{0.4pt}
			$-5.2262r_{p,m}^2+0.0347r_{p,m}^3$~\cite{Pohl:2010zza} & muonic hydrogen 	\\
			\Xhline{0.4pt}
			$-8.084585554\times10^{-7}r_{p,e}^2$ & electronic hydrogen 	\\
			\Xhline{0.4pt}
			$-6.1103r_{d,m}^2$~\cite{Krauth:2015nja,Pohl1:2016xoo} & muonic deuterium 	\\
			\Xhline{0.4pt}
			$-8.091186778\times10^{-7}r_{d,e}^2$ & electronic deuterium 	\\
			\bottomrule\morecmidrules\bottomrule 	
		\end{tabular}
	}
\end{table}
\subsection{Determination of the $r_0$}
As discussed previously, the leading order nuclear size correction to the Lamb shift can be reproduced by assuming that the nucleus is a uniformly charged spheroid. This uniform spheroidal charge distribution is parametrized as
\begin{equation}
\rho(r)=\frac{3}{4\pi r_0^3}\frac1{1+\exp\left(\frac{r-r_0}{a} \right)}.
\end{equation}
We varied the radius $r_0$ of proton and deuteron from $0.1~\mathrm{fm}$ to $4~\mathrm{fm}$ and fitted the energy shifts $\Delta E$~($E_{2S}-E_{2P}$) with the homogeneous binomial of the radius $r_0$. The numerical energy levels are shown in Table~\ref{table:radius-p} for the hydrogen and Table~\ref{table:radius-d} for the deuterium.
\begin{table}[htpb]
	\centering
	\caption{\label{table:radius-p}The energy levels of muonic hydrogen~(left half panel) and electronic hydrogen~(right half panel) with the uniformly charged spheroidal proton.}
	\scalebox{0.8}{
	\begin{tabular}{l*{7}{c}}
		\toprule\morecmidrules\toprule
		\multirow{3}*{$r_0$($\mathrm{fm}$)} & \multicolumn{6}{c}{Energy levels} \\
		\Xcline{2-4}{0.4pt}\Xcline{5-7}{0.4pt}
		& \multicolumn{3}{c}{muonic hydrogen($\mathrm{keV}$)}& \multicolumn{3}{c}{electronic hydrogen($\times10^{-3}\ \mathrm{keV}$)}\\
		\Xcline{2-4}{0.4pt}\Xcline{5-7}{0.4pt}
		& 1S & 2P & 2S& 1S & 2P & 2S \\
		\Xhline{0.8pt}
		0	& -2.528493308  & -0.632123327 & -0.632123327& -13.59828697872  & -3.39957174468 & -3.39957174468\\
		0.2	& -2.528492307  & -0.632123325 & -0.632123199& -13.59828697846  & -3.39957174468 & -3.39957174465\\
		0.4	& -2.528489345  & -0.632123326 & -0.632122829& -13.59828697830  & -3.39957174467 & -3.39957174460\\
		0.6	& -2.528484326  & -0.632123327 & -0.632122203& -13.59828697715  & -3.39957174467 & -3.39957174447\\
		0.8	& -2.528477425  & -0.632123327 & -0.632121341& -13.59828697617  & -3.39957174468 & -3.39957174436\\
		1.0	& -2.528468472  & -0.632123327 & -0.632120222& -13.59828697512  & -3.39957174468 & -3.39957174423\\
		2.0	& -2.528394415  & -0.632123327 & -0.632110965& -13.59828696324  & -3.39957174467 & -3.39957174272\\
		4.0	& -2.528101354  & -0.632123327 & -0.632074330& -13.59828691712  & -3.39957174468 & -3.39957173698\\
		\bottomrule\morecmidrules\bottomrule
	\end{tabular}
	}
\end{table}

\begin{table}[htpb]
	\centering
	\caption{\label{table:radius-d}The energy levels of muonic deuterium~(left half panel) and electronic deuterium~(right half panel) with the uniformly charged spheroidal deuteron.}
	\scalebox{0.8}{
		\begin{tabular}{l*{7}{c}}
			\toprule\morecmidrules\toprule
			\multirow{3}*{$r_0$($\mathrm{fm}$)} & \multicolumn{6}{c}{Energy levels} \\
			\Xcline{2-4}{0.4pt}\Xcline{5-7}{0.4pt}
			& \multicolumn{3}{c}{muonic deuterium($\mathrm{keV}$)}& \multicolumn{3}{c}{electronic deuterium($\times10^{-3}\ \mathrm{keV}$)}\\
			\Xcline{2-4}{0.4pt}\Xcline{5-7}{0.4pt}
			& 1S & 2P & 2S& 1S & 2P & 2S \\
			\Xhline{0.8pt}
			0	& -2.663200422  & -0.665800105 & -0.665800105& -13.60198706144  & -3.40049676536 & -3.40049676536\\
			0.2	& -2.663199252  & -0.665800103 & -0.665799956& -13.60198706122  & -3.40049676536 & -3.40049676532\\
			0.4	& -2.663195790  & -0.665800105 & -0.665799524& -13.60198706102  & -3.40049676536 & -3.40049676530\\
			0.6	& -2.663189930  & -0.665800105 & -0.665798793& -13.60198705987  & -3.40049676533 & -3.40049676513\\
			0.8	& -2.663181867  & -0.665800106 & -0.665797786& -13.60198705884  & -3.40049676535 & -3.40049676502\\
			1.0	& -2.663171406  & -0.665800106 & -0.665796478& -13.60198705788  & -3.40049676536 & -3.40049676491\\
			2.0	& -2.663084922  & -0.665800106 & -0.665785668& -13.60198704598  & -3.40049676536 & -3.40049676342\\
			4.0	& -2.662742880  & -0.665800106 & -0.665742910& -13.60198699982  & -3.40049676536 & -3.40049675765\\
			\bottomrule\morecmidrules\bottomrule
		\end{tabular}
	}
\end{table}

\begin{figure}[htpb]
	\centering
	\includegraphics[width=1.0\textwidth]{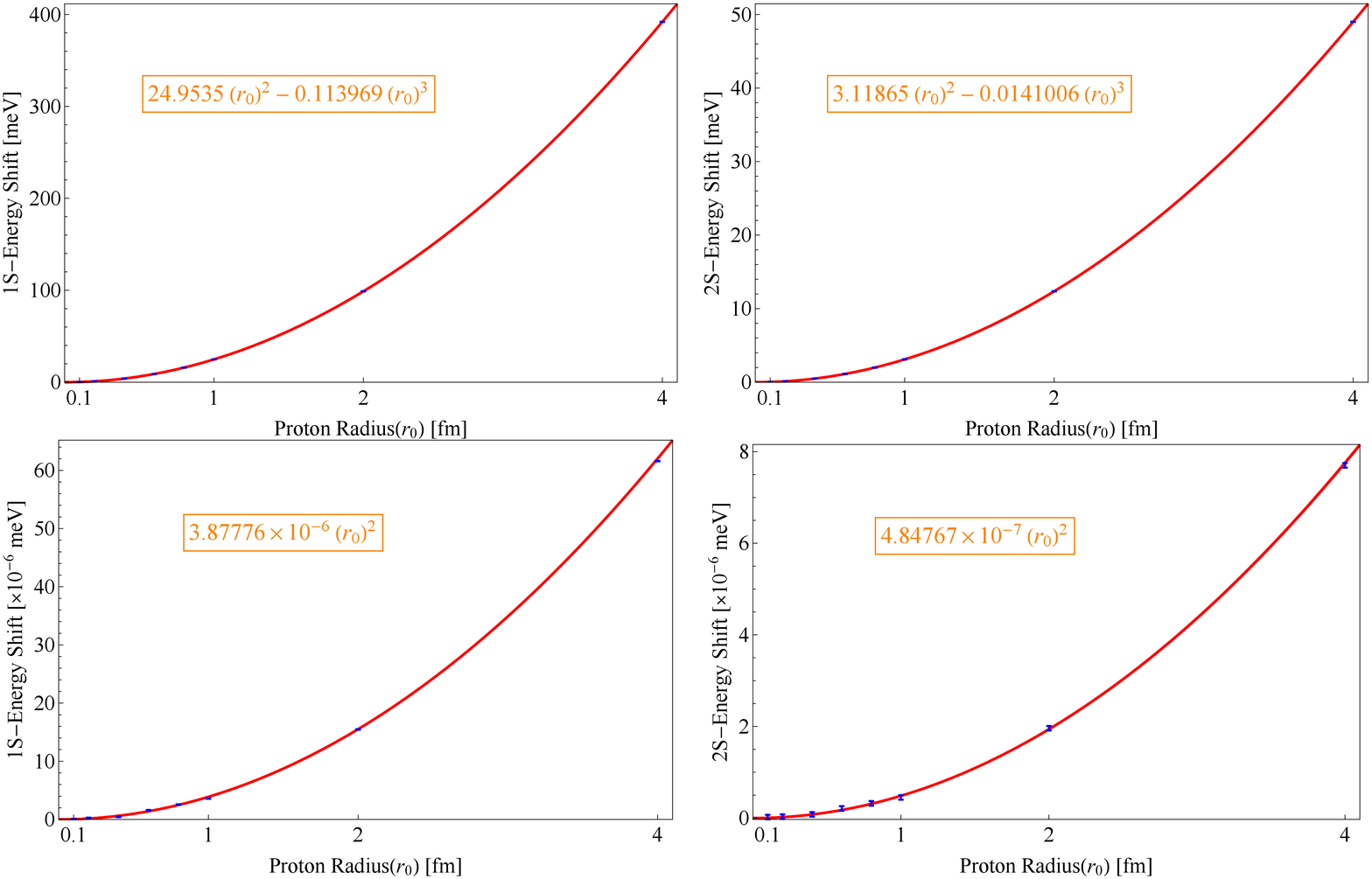}
	\caption{\label{figure:spherical-p}
		Energy shifts of the muonic~(above half) and electronic~(lower half) hydrogen with the radius of spheroidal proton increasing from $0.1~\mathrm{fm}$ to $4.0 ~\mathrm{fm}$: (left) energy level 1S; (right) energy level 2S. The energy shifts are in units of $\mathrm{meV}$. The orange framed shows the best fit.}
\end{figure}

\begin{figure}[htpb]
	\centering
	\includegraphics[width=1.0\textwidth]{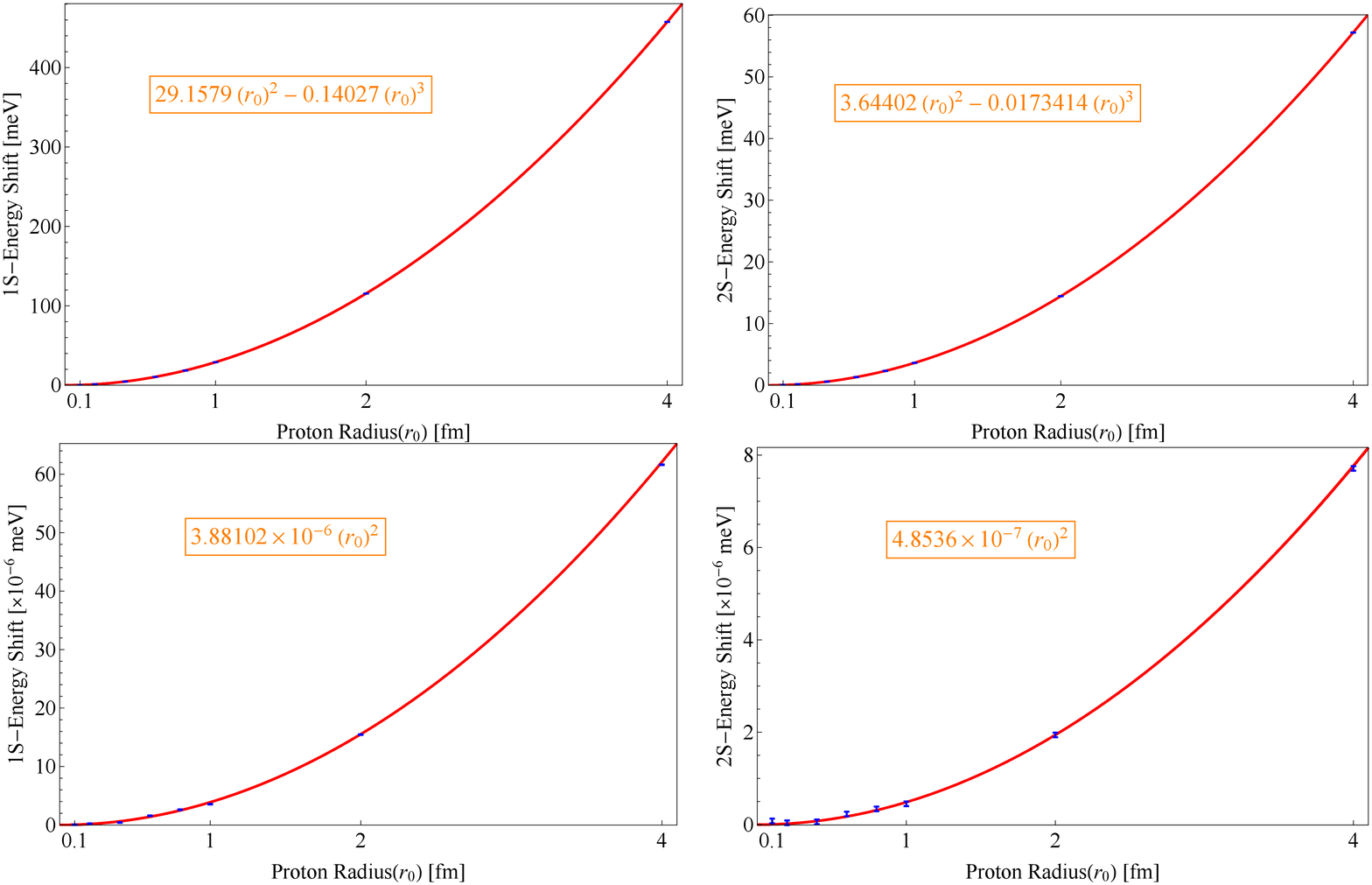}
	\caption{\label{figure:spherical-d}
		Energy shifts of the muonic~(above half) and electronic~(lower half) deuterium with the radius of spheroidal deuteron increasing from $0.1~\mathrm{fm}$ to $4.0 ~\mathrm{fm}$: (left) energy level 1S; (right) energy level 2S. The energy shifts are in units of $\mathrm{meV}$. The orange framed shows the best fit.}
\end{figure}

The energy shifts of the $2P$ states are negligible when the proton charge diffuse isotropically in the space. This property of $2P$ states indicates that the accuracy of our numerical calculation reaches $1.0\times10^{-2}\ \mmev$ for the muonic atoms and $1.0\times10^{-8}\ \mmev$ for the electronic atoms. At this accuracy level, the number of Gaussian basis functions we used, $n_{max}$, equals 44 for the muonic atoms and 70 for the electronic atoms. The fitting results are presented in the Fig.~\ref{figure:spherical-p} for the proton and Fig.~\ref{figure:spherical-d} for the deuteron. The expressions appearing in the orange frames show the best fits. Note that for the energy shifts of muonic atoms, the cubic terms of $r_0$ have to be included to obtain the best fits. Comparing these best fits with the QED formula of the leading nuclear size effect in Eq.~\eqref{eq:finiteness}, we can obtain immediately the effective ratio, that is, the RMS proton charge radius in the theoretical expression of Lamb shift over the effective proton radius in the uniform spheroidal charge distribution assumption $\sqrt{\left\langle r^2\right\rangle}/r_0$. The results are collected in Table~\ref{table:pratio} for the proton and Table~\ref{table:dratio} for the deuteron. It shows the consistent result for the different states as we expected. Considering the present accuracy level and the key points to be issued, we adopt that $\sqrt{\left\langle r^2\right\rangle}/r_0=0.774$ in our following calculation.

\begin{table}[htpb]
	\centering
	\caption{\label{table:pratio}Comparison of the leading nucleon size effects to the $nS$ energy levels of muonic and electronic hydrogen between the QED calculation and the effective spheroidal proton assumption. The digits are the coefficients of the leading nucleon size correction which is proportional to $\left\langle r^2\right\rangle$. These coefficients are in the unit of $\mathrm{meV/fm^2}$.}
	\begin{tabular}{l*{4}{c}}
		\toprule\morecmidrules\toprule
		\multirow{2}*{Comparison} & \multicolumn{2}{c}{$1S$} & \multicolumn{2}{c}{$2S$}\\
		\Xcline{2-3}{0.4pt}\Xcline{4-5}{0.4pt}
		& muonic & electronic($\times 10^{-6}$) & muonic & electronic($\times 10^{-6}$) \\
\Xhline{0.8pt}
QED 	 			& 41.5796  	& 6.46767 	& 5.19745  	& 0.808459\\
Spheroidal Assumption	 		& 24.9533  	& 3.87776 	& 3.11852  	& 0.484767\\
\Xhline{0.8pt}
Effective ratio($\sqrt{\left\langle r_p^2\right\rangle/r_0^2}$) 	& 0.774683	& 0.774313	& 0.774603  	& 0.774350\\
		\bottomrule\morecmidrules\bottomrule
	\end{tabular}
\end{table}
\begin{table}[htpb]
	\centering
	\caption{\label{table:dratio}Comparison of the leading nucleon size effects to the $nS$ energy levels of muonic and electronic deuterium between the QED calculation and the effective spheroidal proton assumption. The digits are the coefficients of the leading nucleon size correction which is proportional to $\left\langle r^2\right\rangle$. These coefficients are in the unit of $\mathrm{meV/fm^2}$.}
	\begin{tabular}{l*{4}{c}}
		\toprule\morecmidrules\toprule
		\multirow{2}*{Comparison} & \multicolumn{2}{c}{$1S$} & \multicolumn{2}{c}{$2S$}\\
		\Xcline{2-3}{0.4pt}\Xcline{4-5}{0.4pt}
		& muonic & electronic($\times 10^{-6}$) & muonic & electronic($\times 10^{-6}$) \\
		\Xhline{0.8pt}
		QED 	 			& 48.5855  	& 6.47295 	& 6.07319  	& 0.809119\\
		Spheroidal Assumption	 		& 29.1582  	& 3.88102 	& 3.64425  	& 0.485360\\
		\Xhline{0.8pt}
		Effective ratio($\sqrt{\left\langle r_d^2\right\rangle/r_0^2}$) 	& 0.774688	& 0.774322	& 0.774632  	& 0.774508\\
		\bottomrule\morecmidrules\bottomrule
	\end{tabular}
\end{table}

\subsection{Calculations with the prolate assumption}
After obtaining the relationship between the effective charge radius appeared in the charge distribution assumption and the extracted charge radius, we can estimate the energy shifts of hydrogen or deuterium when the spheroidal proton or deuteron with a given radius $r_p$ or $r_d$ is pressed into a prolate shape. Before going forward, let us illustrate firstly the strategy with more details. At present, the proton charge radius extracted from the electronic hydrogen spectroscopy data, labeled as $r_{p,e;exp}$, is recognized typically as $0.875~\mathrm{fm}$. And similarly, $r_{p,m;exp}$ is accepted as $0.842~\mathrm{fm}$. That is,
\begin{align}\label{eq:explambshift}
\Delta E_{p,m;exp}^{LS}&=\Delta E_{p,m;0}-5.2262r_{p,m;exp}^{2}+0.0347r_{p,m;exp}^3,\notag\\
\Delta E_{p,e;exp}^{LS}&=\Delta E_{p,e;0}-8.084585554\times10^{-7}r_{p,e;exp}^2,
\end{align}
where $\Delta E_{p,m;0}$ and $\Delta E_{p,e;0}$ denote the radius-independent terms from QED calculation in the Lamb shift. $\Delta E_{p,m;exp}^{LS}$ and $\Delta E_{p,e;exp}^{LS}$ denote the $2S-2P$ Lamb shifts of muonic and electronic hydrogen measured in the experiments. If we consider the axial deformed nucleus, there must be an additional contribution, denoted as $\Delta E_{def}$, to the right side in Eq.~\eqref{eq:explambshift}. Note that it is a function of the effective radius $r_0$ and the deformation parameter $\beta$. Then the corrected theoretical expression for the lamb shifts can be rewritten as
\begin{align}\label{eq:deflambshift}
\Delta E_{p,m;exp}^{LS}&=\Delta E_{p,m;0}-5.2262r_{p,m}^{2}+0.0347r_{p,m}^{3}+\Delta E_{p,m;def}(r_{p,m},\beta),\notag\\
\Delta E_{p,e;exp}^{LS}&=\Delta E_{p,e;0}-8.084585554\times10^{-7}r_{p,e}^{2}+\Delta E_{p,e;def}(r_{p,e},\beta).
\end{align}
Where $r_p=0.774 r_0$. Combing Eq.~\eqref{eq:explambshift} and Eq.~\eqref{eq:deflambshift}, we can obtain
\begin{align}\label{eq:defconditionp}
\Delta E_{p,m;def}(r_{p,m},\beta_p)&=5.2262\left(r_{p,m}^{2}-r_{p,m;exp}^{2}\right)-0.0347\left(r_{p,m}^{3}-r_{p,m;exp}^{3}\right),\notag\\
\Delta E_{p,e;def}(r_{p,e},\beta_p)&=8.084585554\times10^{-7}\left(r_{p,e}^{2}-r_{p,e;exp}^2\right).
\end{align}
Note that the symbol $\beta_p$ is used for the proton to distinguish with deuteron in the following. Eq.~\eqref{eq:defconditionp} states the correlations between the effective radius $r_0$~($r_p/0.774$) and the deformation parameter $\beta$ for the muonic and electronic hydrogen respectively. And if the deformation of proton is exactly the source of the proton radius puzzle, the values of experimental extracted radius from electronic and muonic hydrogen can get consistent with each other in the case that inner structure of proton in the electronic and muonic hydrogen can be parameterized with a same set of ($r_{p}$, $\beta_p$), that is, $r_{p,m}=r_{p,e}\equiv r_p$. Therefore the crucial part of our work is to find a set of solution($r_p,\beta_p$) for the equation group Eq.~\eqref{eq:defconditionp}. For the deuterium case, the strategy is similar and the constraint equations can be read as
\begin{align}\label{eq:defconditiond}
\Delta E_{d,m;def}(r_{d,m},\beta_d)&=6.1103\left(r_{d,m}^{2}-r_{d,m;exp}^{2}\right)\notag\\
\Delta E_{d,e;def}(r_{d,e},\beta_d)&=8.091186778\times10^{-7}\left(r_{d,e}^{2}-r_{d,e;exp}^2\right).
\end{align}
Here $r_{d,e;exp}$ denotes the deuteron charge radius extracted from the electronic deuterium spectroscopy data, which is recognized typically as $2.1424~\mathrm{fm}$ and $r_{d,m;exp}$ is accepted as $2.12562~\mathrm{fm}$.

In practice, equation groups Eq.~\eqref{eq:defconditionp} and Eq.~\eqref{eq:defconditiond} can not be solved straightforwardly since the functional expression of the additional contribution $\Delta E_{def}$ about the variable $r_p$($r_d$) and $\beta_p$($\beta_d$) is unknown. Alternatively, we calculated the electromagnetic potential derived by the uniformly prolate charge distribution numerically and solved the energy levels by means of the variational method. After subtracting the energy shifts between $2S$ and $2P$ states caused by the uniformly spheroidal nucleus with the same effective radius $r_p$($r_d$), we can obtain the value of $\Delta E_{def}$ for a given set of parameter $r_p$ and $\beta_p$. To find the solutions of equation groups Eq.~\eqref{eq:defconditionp} and Eq.~\eqref{eq:defconditiond}, we calculate a series of $\Delta E_{p,m;def}$ corresponding to the different $\beta_p$ in the range of $[0,~0.6]$ with a fixed effective radius $r_p$ for the muonic atoms. Then with $r_{p,m;exp}$ given as an input, we can solve the muonic equations in equation groups Eq.~\eqref{eq:defconditionp} to obtain the $\beta_p$ by using the fitting polynomial formula of $\Delta E_{p,m;def}$ as a function of $\beta_p$. With the same parameter ($r_p$, $\beta_p$), we can obtain the value of $\Delta E_{p,e;def}$ for the electronic atoms and solve the electronic equations in equation groups Eq.~\eqref{eq:defconditionp} to obtain the $r_{p,e;exp}^\prime$. Finally, if the derived $r_{p,e;exp}^\prime$ equals the inputted $r_{p,m;exp}$, the set of parameter ($r_p$, $\beta_p$) is the solution of equation group Eq.~\eqref{eq:defconditionp}. The case is similar for the deuteron.

Unfortunately, from our calculation it is found that the nonspherical correction to Lamb shifts $\Delta E_{def}$ is negative for both of electronic and muonic hydrogen and also for the electronic and muonic deuterium. It implies that the deformation of proton will enlarge the energy shifts between $2S$ and $2P$ states which are caused by the finite size of nucleus. As equation groups Eq.~\eqref{eq:defconditionp} and Eq.~\eqref{eq:defconditiond} shown, the negative contribution $\Delta E_{def}$ calls for the smaller inner size of proton $r_p$ compared to the accepted value $r_{p,exp}$. Actually, the solution of equation groups Eq.~\eqref{eq:defconditionp} and Eq.~\eqref{eq:defconditiond} cannot locate in the small radius region since it is expected that the absolute magnitude of radius shift ($r_{p,m}-r_{p,m;exp}$) in the muonic system should be larger than that one in the electronic atoms and the reported value $r_{p,m;exp}$ is smaller than the previous accepted $r_{p,e;exp}$. 
As a result, the deformation of proton would lead to that the charge radius extracted from the electronic measurement should be smaller instead of larger than the corresponding one in the muonic measurement which assumed uniform spheroidal charge distribution. In fact, there is another term of the Lamb shifts to be corrected by the deformation of nucleus. It is the $\order{(Z\alpha)^5}$ nuclear polarizability contribution which is proportional to the electric polarizability of proton and its sign is opposite with the leading finite size contribution~\cite{Eides:2000xc}. The electric polarizability is the fundamental structure constant of proton which will be modified in the deformed assumption. However, the magnitude of such correction have been constrained strictly by the proton Compton scattering experiments which have determined the electric polarizability of proton with small uncertainty~\cite{MacGibbon:1995in}. The possibility that the correction from this term can be commensurate with $0.31\ \mmev$ is excluded. Thereby the situation seems to be clear. The deformation of the proton as well as other known theoretical attempts cannot explain the contradiction between the extracted proton radius from the muonic hydrogen atom experiment and the larger average proton radius extracted from previous electron-proton experiments. The solution of the proton radius puzzle may come from experimental side as suggested by the recent proton charge radius measurements of the electronic hydrogen spectroscopy reported in Refs.~\cite{Beyer:2017gug,Bezginov:2019mdi}. And also the analysis of the electron-proton scattering data by implementing the theoretical constrains on the nucleon form factors from the dispersion relation point of view shows the smaller proton charge radius in Refs.~\cite{Lorenz:2012tm,Lorenz:2014vha,Lorenz:2014yda}. In the following part, we will illustrate how the deformation of nucleus affects the extracted charge radius from experimental Lamb shifts.

In order to estimate the behavior of extracted charge radius under the change of deformation parameter $\beta$, we choose the preciser charge radius which is extracted from the muonic measurement $r_{p,m;exp}$ as the input constant and solve the related extracted radius from electronic system $r_{p,e;exp}^\prime$ for a given $\beta$ from the equation group Eq.~\eqref{eq:defconditionp}. In such way, we can estimate the preferred nucleus charge radius for the electronic measurement in the deformed scenario. Note that this strategy is based on the assumption that if the nucleus has no non-spherical defomation, the extracted charge radius from the electronic measurement should be in good agreement with the muonic measurement, that is, $r_{p,e;exp}^\prime=r_{p,m;exp}$ for $\beta=0$. The numerical energy intervals are collected in Table~\ref{table:beta-p} for the electronic and muonic hydrogen and Table~\ref{table:beta-d} for the electronic and muonic deuterium respectively. Note that we varied the $r_p$ in the range of $[0.78,~0.83]~\mathrm{fm}$ for the proton and $[2.06,~2.11]~\mathrm{fm}$ for the deuteron and the deformation parameter $\beta$ is varied in the range $[0,~0.6]$ for both the proton and deuteron. Here 0.6 refs to the suggestion in Ref.~\cite{Buchmann:2001gj}.
\begin{table}[htpb]
	\centering
	\caption{\label{table:beta-p}The energy intervals between $2S$ and $2P$ states of muonic and electronic hydrogen with the uniformly charged prolate proton. Here $r_p$ varies in the range of $[0.78,~0.83]~\mathrm{fm}$. The density parameter $\beta$ changes from $0.0$ to $0.6$ and $r_0=r_p/0.774$.}
	\begin{tabular}{l*{8}{c}}
		\toprule\morecmidrules\toprule
		\multirow{2}*{$r_p$($\mathrm{fm}$)} & \multicolumn{7}{c}{Muonic hydrogen $\Delta E$ ($\mathrm{meV}$)} \\
		\Xcline{2-8}{0.4pt}
		& $\beta=0$ & $\beta=0.1$ & $\beta=0.2$ & $\beta=0.3$ & $\beta=0.4$ & $\beta=0.5$ & $\beta=0.6$ \\
		\Xhline{0.8pt}
		0.78	& 3.154& 3.244& 3.441& 3.765& 4.241& 4.906& 5.815\\
		0.79	& 3.236& 3.327& 3.530& 3.862& 4.350& 5.033& 5.966\\
		0.80	& 3.319& 3.413& 3.620& 3.960& 4.460& 5.161& 6.117\\
		0.81	& 3.402& 3.497& 3.710& 4.059& 4.571& 5.290& 6.270\\
		0.82	& 3.486& 3.585& 3.803& 4.160& 4.685& 5.420& 6.425\\
		0.83	& 3.570& 3.672& 3.897& 4.262& 4.800& 5.553& 6.583\\
		\midrule
		\multirow{2}*{$r_p$($\mathrm{fm}$)} & \multicolumn{7}{c}{Electronic hydrogen $\Delta E$ ($\times10^{-7}\ \mathrm{meV}$)}  \\
		\Xcline{2-8}{0.8pt}
		& $\beta=0$ & $\beta=0.1$ & $\beta=0.2$ & $\beta=0.3$ & $\beta=0.4$ & $\beta=0.5$ & $\beta=0.6$   \\
		\Xhline{0.8pt}
		0.78	& 4.6& 4.8& 5.0& 5.4& 6.1& 7.2& 8.5\\
		0.79	& 4.7& 4.8& 5.1& 5.6& 6.5& 7.3& 8.8\\
		0.80	& 4.9& 5.0& 5.3& 5.7& 6.5& 7.6& 8.9\\
		0.81	& 5.0& 5.1& 5.5& 6.0& 6.8& 7.7& 9.2\\
		0.82	& 5.2& 5.3& 5.5& 6.2& 6.8& 8.0& 9.5\\
		0.83	& 5.3& 5.4& 5.8& 6.3& 7.2& 8.2& 9.8\\
		\bottomrule\morecmidrules\bottomrule
	\end{tabular}
\end{table}

\begin{figure}[htpb]
	\centering
	\includegraphics[width=1.0\textwidth]{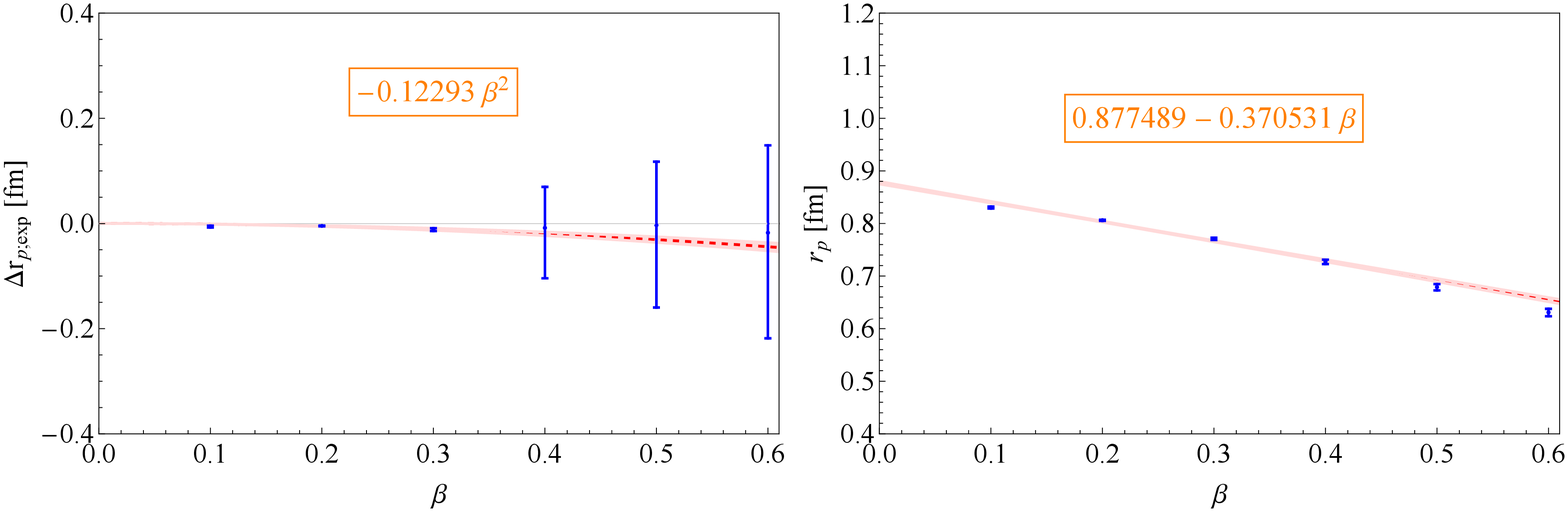}
	\caption{\label{figure:proton}
		The fitting results for the proton: deviation between the proton charge radius extracting from the electronic and muonic hydrogen $\Delta r_{p,exp}$ vs $\beta$~(left panel), inner size of the proton $r_p$ vs $\beta$~(right panel). The orange framed shows the best fit. $\Delta r_{p,exp}\equiv r_{p,e;exp}^\prime-r_{p,m;exp}$.}
\end{figure}

\begin{figure}[htpb]
	\centering
	\includegraphics[width=1.0\textwidth]{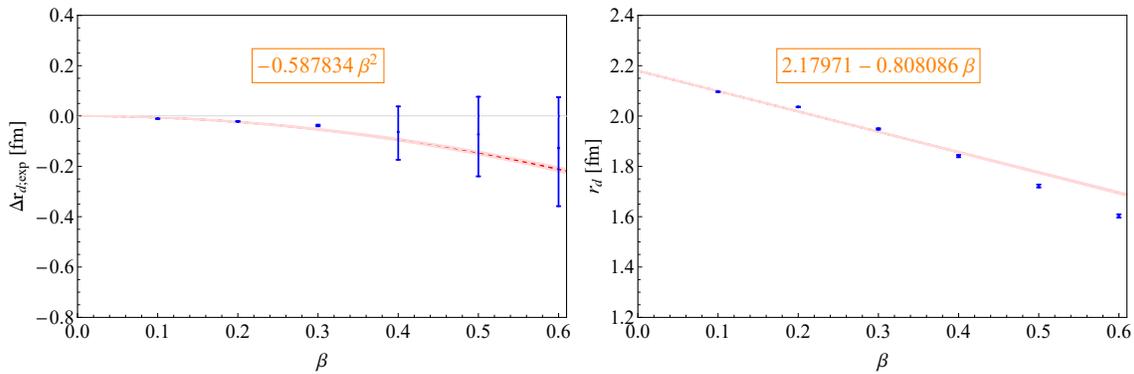}
	\caption{\label{figure:deuteron}
		The fitting results for the deuteron: deviation between the deuteron charge radius extracting from the electronic and muonic deuterium $\Delta r_{d,exp}$ vs $\beta$~(left panel), inner size of the deuteron $r_d$ vs $\beta$~(right panel). The orange framed shows the best fit. $\Delta r_{d,exp}\equiv r_{d,e;exp}^\prime-r_{d,m;exp}$.}
\end{figure}

\begin{table}[htpb]
	\centering
	\caption{\label{table:beta-d}The energy intervals between $2S$ and $2P$ states of muonic and electronic deuterium with the uniformly charged prolate deuteron. Here $r_d$ varies in the range of $[2.06,~2.11]~\mathrm{fm}$. The density parameter $\beta$ changes from $0.0$ to $0.6$ and $r_0=r_p/0.774$.}
	\begin{tabular}{l*{8}{c}}
		\toprule\morecmidrules\toprule
		\multirow{2}*{$r_d$($\mathrm{fm}$)} & \multicolumn{7}{c}{Muonic deuterium $\Delta E$ ($\mathrm{meV}$)} \\
		\Xcline{2-8}{0.4pt}
		& $\beta=0$ & $\beta=0.1$ & $\beta=0.2$ & $\beta=0.3$ & $\beta=0.4$ & $\beta=0.5$ & $\beta=0.6$ \\
		\Xhline{0.8pt}
		2.06	& 25.487& 26.218& 27.819& 30.430& 34.263& 39.624& 46.950\\
		2.07	& 25.733& 26.471& 28.089& 30.725& 34.595& 40.007& 47.404\\
		2.08	& 25.981& 26.726& 28.358& 31.020& 34.927& 40.391& 47.860\\
		2.09	& 26.230& 26.982& 28.630& 31.317& 35.262& 40.780& 48.317\\
		2.10	& 26.480& 27.239& 28.903& 31.616& 35.598& 41.168& 48.778\\
		2.11	& 26.731& 27.497& 29.178& 31.915& 35.935& 51.555& 49.240\\
		\midrule
		\multirow{2}*{$r_d$($\mathrm{fm}$)} & \multicolumn{7}{c}{Electronic deuterium $\Delta E$ ($\times10^{-7}\ \mathrm{meV}$)}\\ \Xcline{2-8}{0.4pt}
		& $\beta=0$ & $\beta=0.1$ & $\beta=0.2$ & $\beta=0.3$ & $\beta=0.4$ & $\beta=0.5$ & $\beta=0.6$  \\
		\Xhline{0.8pt}
		2.06	& 34.1& 34.6& 36.6& 39.6& 44.3& 50.9& 59.7\\
		2.07	& 34.4& 35.0& 36.8& 40.0& 44.6& 51.3& 60.4\\
		2.08	& 34.8& 35.3& 37.1& 40.3& 45.3& 51.8& 61.0\\
		2.09	& 34.9& 35.6& 37.6& 40.7& 45.7& 52.5& 61.5\\
		2.10	& 35.3& 35.9& 37.8& 41.1& 46.1& 52.8& 62.1\\
		2.11	& 35.6& 36.3& 38.3& 41.5& 46.4& 53.3& 62.8\\
		\bottomrule\morecmidrules\bottomrule
	\end{tabular}
\end{table}
According to the numerical results, the correction to the leading finite size contribution caused by the deformation of nucleus $\Delta E_{def}$ is approximately a linear function of the effective radius $r_p$ for a given deformation parameter $\beta$ both in the proton and deuteron systems. And the deviation of the extracted charge radius from the electronic and muonic measurement $\Delta r_{p(d),exp}$~($\equiv r_{p(d),e;exp}^\prime - r_{p(d),m;exp}$) can be expressed roughly as the quadratic formula of $\beta$. The fitting results are presented in Fig.~\ref{figure:proton} for the proton and Fig.~\ref{figure:deuteron} for the deuteron.

If we adopt the central value $0.833~\mathrm{fm}$ in Ref.~\cite{Bezginov:2019mdi} as the experimental extracted proton charge radius $r_{p,e;exp}^\prime$, the deformation parameter $\beta$ can be estimated as $0.245_{-0.245}^{+0.169}$ and correspondingly the inner size of proton can be obtain as $0.7867_{-0.0648}^{+0.0931}~\mathrm{fm}$. From the deformation parametrization of Eq.~\eqref{eq:surface}, the center value of the inner structure corresponds to the three-dimensional prolate proton with the 0.9083-$\mathrm{fm}$ long axis and 0.7259-$\mathrm{fm}$ short axis. Note that our extracted central value for the deformation parameter $\beta=0.245$ is smaller than that of Ref.~\cite{Buchmann:2001gj}. In that paper, authors suggested that proton has the prolate deformation and claimed a ratio of major to minor semi-axes around 1.73 which is equivalent to $\beta=0.62$. As for the deuteron, its D-state probability has been proposed in early work~\cite{Feshbach:1951zz} and has been a classic problem of nuclear physic. The value of the deuteron D-state probability has been recognized typically in the range of $2\%\sim 7\%$~\cite{Kelkar:2016qdt,Zhaba:2017syr,Koohrokhi:2018tek}. Actually, the non-vanishing D-wave component in the deuteron wave function indicates the existence of pure quadrupole deformation for deuteron. Refocusing on our parametrization of charge distribution in Eq.~\eqref{eq:surface}, the deuteron D-state probability $P_D$ can be translated to the deformation parameter $\beta$,
\begin{equation}
P_D=\sqrt{\left(\frac\beta{2\sqrt{\pi}}\right)^2/\left(\left(\frac\beta{2\sqrt{\pi}}\right)^2+1\right)}.
\end{equation}
Thereby the deformation parameter $\beta$ for the deuteron can be estimated to be in the range of  $0.0709\sim 0.2488$. As shown by the fits in Fig.~\ref{figure:deuteron}, the corresponding extracted radius deviation is in the range of $-(0.002845\sim 0.03774)~\mathrm{fm}$ and the inner size of deuteron $r_d$ is in the range of $1.9777\sim 2.1230~\mathrm{fm}$. Then it is expected that the deuteron charge radius extracting from the electronic measurement should be in the range of $(2.0879\sim 2.1228)~\mathrm{fm}$. For the upper-limit of D-state probability, that is $\beta=0.2488$, the 3-dimensional structure of deuteron is described as a prolate with the long axis of $2.2873~\mathrm{fm}$ and the short axis of $1.8219~\mathrm{fm}$. Although our numerical calculation bears large uncertainty, the effect on the extracted nucleus charge radius from  nuclear deformation has been clearly demonstrated and it will help us to understand the deformation shape of proton and deuteron.

\section{Summary} \label{sec:summary}
Proton radius puzzle has been a fundamental physical problem since the precise proton charge radius extracted from the muonic hydrogen was reported in 2010. In the present work, we assumed the proton has pure quadrupole deformation and calculated the energy shifts of electronic and muonic hydrogen by means of the Gaussian Expansion Method. It is found that the deformation of proton will render the charge radius extracting from the electronic spectroscopy $r_{p,e;exp}$ smaller than that of the muonic measurement $r_{p,m;exp}$. If the central value of the latest electronic measurement~\cite{Bezginov:2019mdi} is adapted, it would correspond to the prolate proton with $r_p=0.787~\mathrm{fm}$ and $\beta=0.245$, that is, the 3-dimensional charge distribution with the 0.9083-$\mathrm{fm}$ long axis along the z axis and two 0.7259-$\mathrm{fm}$ short axes along x, y-axis directions. Besides, the discussion on the deuteron is also included in our work. The numerical calculation suggests that the $7\%$ D-state probability is related to the prolate deuteron with $r_d=1.978~\mathrm{fm}$ and $\beta=0.249$, that is, the 3-dimensional charge distribution with the 2.287-$\mathrm{fm}$ long axis along the z axis and two 1.822-$\mathrm{fm}$ short axes along x, y-axis directions. And accordingly, the deuteron charge radius extracting from the electronic measurement is expected to be $2.089~\mathrm{fm}$. To draw the definite conclusions on the deformation shape of the proton and deuteron, it needs more precise experiments both in electronic and muonic sectors in the future. This study of nuclear deformation from the corresponding electronic and muonic atomic transitions can be easily extended to other nuclei.   

\bigskip

\section*{Acknowledgments}

We thank Shangui Zhou and Xiangxiang Sun for useful discussions. This project is supported by NSFC under Grant
No. 11621131001 (CRC110 cofunded by DFG and NSFC) and Grant No. 11835015.



\end{document}